\documentclass[preprint,12pt,sort&compress]{elsarticle}

\usepackage{amsmath,amssymb}
\usepackage{booktabs}
\usepackage{graphicx}
\usepackage{microtype}
\usepackage{float}
\usepackage{tabularx}
\usepackage{threeparttable}
\usepackage{array}
\usepackage{xurl}
\usepackage{hyperref}
\hypersetup{hidelinks}

\journal{Biomedical Signal Processing and Control}

\begin{document}
\begin{frontmatter}

\title{Cross-Sectional Separability versus Longitudinal Response in Short-Record Parkinsonian Gait Analysis Using FEG-Pro: Nordic Walking and Adapted Physical Activity}

\author[aff1]{Xuanbao Xiang}
\ead{1184794731@qq.com}
\author[aff2]{Andrei Velichko}
\ead{velichkogf@gmail.com}
\author[aff1]{Xiaobo Rao}
\ead{rxbaizxp@163.com}
\author[aff1]{Jianshe Gao\corref{cor1}}
\ead{gao_jianshe@zzu.edu.cn}
\cortext[cor1]{Corresponding author.}

\affiliation[aff1]{organization={School of Mechanical and Power Engineering, Zhengzhou University},
  addressline={Science Avenue 100}, city={Zhengzhou}, postcode={450001}, state={Henan}, country={China}}
\affiliation[aff2]{organization={Institute of Physics and Technology, Petrozavodsk State University},
  city={Petrozavodsk}, postcode={185910}, country={Russia}}

\begin{abstract}
\textbf{Objective:} Machine-learning separation of rehabilitation cohorts does not inherently establish a differential intervention response. This study used Forecast-Error Growth Profiling (FEG-Pro) to distinguish cross-sectional cohort separability from subject-level longitudinal change following Nordic Walking (NW) and Adapted Physical Activity (APA) in Parkinson's disease.

\textbf{Methods:} Publicly available short gait records from 24 participants (NW=14, APA=10) were analyzed. Lower-limb signals at baseline and 12 weeks were transformed into FEG-Pro and Forecast-Error Distribution Entropy descriptors. We compared individual change scores ($\Delta=T1-T0$) between groups using baseline-adjusted sensitivity analyses and false-discovery-rate (FDR) correction. Additionally, a fully nested machine-learning pipeline evaluated whether multidimensional change vectors could identify the intervention.

\textbf{Results:} The self-selected cohorts already differed clinically at baseline. Among 1,098 extracted features, none exhibited robust between-group differences in longitudinal change after FDR correction or baseline adjustment. Furthermore, nested classification based on multidimensional change vectors failed to perform above chance (mean MCC = $-0.178 \pm 0.228$). In contrast, exploratory cross-sectional models separated the cohorts with peak MCC values of 0.604 at baseline and 0.554 post-intervention.

\textbf{Conclusion:} This cohort did not provide robust evidence of modality-specific longitudinal responses to NW or APA. These findings demonstrate that cross-sectional separability of self-selected cohorts must not be interpreted as an intervention effect; true rehabilitation biomarkers require subject-level longitudinal validation, baseline adjustment, and leakage-safe evaluation.
\end{abstract}


\end{frontmatter}

\section{Introduction}
\label{sec:intro}

Gait impairment is a central manifestation of Parkinson's disease (PD), encompassing reduced stride length and speed, impaired rhythm and automaticity, increased variability and asymmetry, postural instability, freezing of gait, and elevated fall risk \cite{ref1,ref2,ref3,ref4,ref5}. Exercise and physical rehabilitation can improve several of these domains, although the magnitude and specificity of benefit vary across interventions and patient subgroups \cite{ref6,ref7,ref8,ref9}. Instrumented studies further show that rehabilitation-related change may occur in joint kinematics, balance, dual-task performance, cue-dependent coordination, muscle activation, and neural-control correlates, rather than only in macroscopic walking speed \cite{ref10,ref11,ref12,ref13,ref14,ref15,ref16,ref17,ref18,ref19,ref20,ref21,ref22,ref23,ref24,ref25,ref26,ref27}.

Nonlinear descriptors are attractive for detecting subtle changes in dynamic stability and predictability that conventional spatiotemporal variables may overlook. Their clinical use is nevertheless constrained by short recordings: classical largest-Lyapunov-exponent and related approaches may require substantially longer and more stationary sequences than are feasible in frail or mobility-limited cohorts \cite{ref28}. Forecast-Error Growth Profiling (FEG-Pro) was recently introduced to characterize finite-horizon instability from short scalar time series by combining the geometry of forecast-error growth with the entropy of signed forecast-error distributions \cite{ref29}. This produces slope, curvature, fit-quality, reliability, roughness, monotonicity, and FEDE descriptors from relatively short records.

A separate methodological problem concerns the target of machine-learning analysis. In rehabilitation studies, a classifier that separates two cohorts at baseline or after treatment measures \emph{cohort separability}. It does not necessarily measure a \emph{differential intervention response}. This distinction is especially important when participants self-select an intervention and the groups are not randomized. Baseline imbalance, personal preferences, physical capacity, and latent motor phenotype can all support classification before any treatment is delivered. A clinically meaningful comparison of treatment-associated responses should therefore prioritize individual pre--post changes and should test whether those changes differ between groups.

The present study revisits a public PD gait dataset in which participants self-selected Nordic Walking (NW) or Adapted Physical Activity (APA) \cite{ref30}. The objectives were: (1) to determine whether FEG-Pro descriptors computed from short gait records show different subject-level longitudinal changes after NW and APA; (2) to assess whether a leakage-safe machine-learning model can identify the intervention from multidimensional change vectors; and (3) to contrast these longitudinal results with exploratory cross-sectional cohort classification. The central hypothesis was not that cross-sectional separability itself indicates treatment efficacy, but that a valid modality-specific response must be supported by reproducible differences in individual change.

\section{Materials and methods}
\label{sec:methods}

\subsection{Dataset and participants}

Data were obtained from the public dataset ``Dataset on Gait Analysis of Parkinsonian Subjects: Effect of Nordic Walking'' \cite{ref30}. The cohort comprised 24 participants with confirmed PD (Hoehn--Yahr stages 1--3; 19 men and 5 women; age $74.92\pm6.53$ years; body mass $74.60\pm12.37$ kg). Eligibility required stable antiparkinsonian medication, independent walking for more than 200 m during the 6-Minute Walk Test (6MWT), and an Addenbrooke's Cognitive Examination III score of at least 71. Participants voluntarily selected NW ($n=14$) or APA ($n=10$); allocation was not randomized.

The NW program lasted 12 weeks and comprised two 90-min sessions per week emphasizing coordinated upper- and lower-limb movement with poles. APA was also delivered for 12 weeks, with 2--3 sessions of approximately 60 min per week including stretching, balance, coordination, aerobic, and low-intensity resistance activities. Baseline characteristics are summarized in Table~\ref{tab:baseline}. The source study reported nominal baseline differences in 6MWT, Berg Balance Scale (BBS), and walking speed, and the machine-readable gait data confirmed a higher baseline walking speed in NW than APA. These imbalances reinforce the need to distinguish baseline cohort characteristics from intervention-associated change.

\begin{table}[H]
\centering
\caption{Selected baseline characteristics of the self-selected cohorts.}
\label{tab:baseline}
\begin{threeparttable}
\small
\begin{tabular}{lccc}
\toprule
Variable & NW ($n=14$) & APA ($n=10$) & $p$ \\
\midrule
Age, years & $73.64\pm5.93$ & $76.70\pm7.21$ & 0.286 \\
Women/men & 2/12 & 3/7 & 0.615 \\
Body mass, kg & $75.93\pm9.44$ & $72.40\pm15.95$ & 0.542 \\
Height, m & $1.696\pm0.054$ & $1.645\pm0.091$ & 0.138 \\
BMI, kg/m$^2$ & $26.44\pm3.21$ & $26.51\pm3.94$ & 0.961 \\
6MWT & not available in public table & not available in public table & 0.038 \\
BBS & not available in public table & not available in public table & 0.045 \\
Walking speed, m/s & $0.960\pm0.214$ & $0.794\pm0.146$ & 0.034 \\
Stride length, m & $1.049\pm0.195$ & $0.937\pm0.160$ & 0.139 \\
Gait Deviation Index & $87.31\pm5.42$ & $87.16\pm8.32$ & 0.962 \\
\bottomrule
\end{tabular}
\begin{tablenotes}[flushleft]\footnotesize
\item Continuous values are mean $\pm$ standard deviation. The 6MWT and BBS $p$-values are reproduced from the source study; cohort-level values were not available in the supplied machine-readable table. NW, Nordic Walking; APA, Adapted Physical Activity; 6MWT, 6-Minute Walk Test; BBS, Berg Balance Scale.
\end{tablenotes}
\end{threeparttable}
\end{table}

\subsection{Gait acquisition and analyzed signals}

Participants underwent instrumented gait analysis approximately one week before the intervention (T0) and one week after completion (T1), during the medication ON state. Kinematics were collected at 100 Hz using eight Vicon Vero cameras and a full-body marker protocol. Participants walked barefoot at self-selected speed over approximately 6 m. At least five walking trials were acquired and three representative trials were included in the public dataset \cite{ref30,ref31}.

The present analysis used scalar time series from the left and right sides: sagittal-plane hip, knee, and ankle angles and vertical heel displacement. No additional frequency filtering or internal resampling was applied. Each trial therefore remained a short, naturally varying clinical record.

\subsection{FEG-Pro and FEDE feature extraction}

\begin{figure}[H]
\centering
\includegraphics[width=\textwidth]{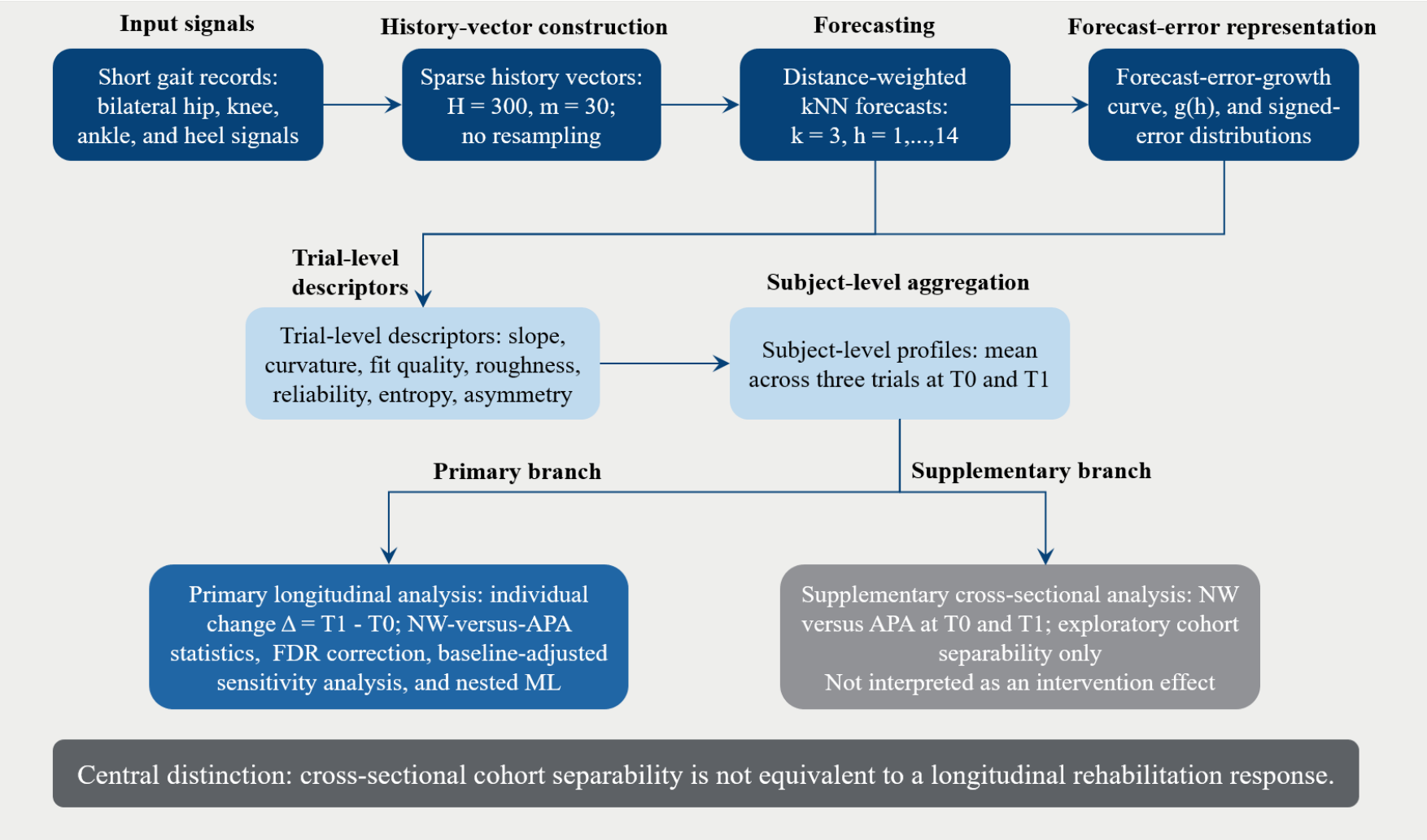}
\caption{FEG-Pro feature extraction and the distinction between the primary longitudinal analysis and supplementary cross-sectional cohort classification. The central inferential target was the individual change $\Delta=T1-T0$, not group membership at a single time point.}
\label{fig:workflow}
\end{figure}

Figure~\ref{fig:workflow} summarizes the analytical pipeline. For a scalar series $x_0,\ldots,x_{N-1}$, FEG-Pro constructs a sparse history vector spanning $H$ samples,
\begin{equation}
\mathbf{s}_i=\left[x_{i+o_1},x_{i+o_2},\ldots,x_{i+o_m}\right],
\end{equation}
where $m=30$ and the integer offsets $o_j$ are approximately equally spaced from 0 to $H-1$. In the fixed configuration used here, $H=300$, $m=30$, and the history coordinates were standardized using training-set statistics. A distance-weighted $k$-nearest-neighbor regressor ($k=3$) was fitted to the first 60\% of the temporally ordered history vectors and evaluated on the remaining 40\%.

For forecast horizon $h=1,\ldots,14$, the signed prediction error was
\begin{equation}
\eta_i(h)=x_{i+H+h-1}-\widehat{x}_{i+H+h-1}.
\end{equation}
The forecast-error-growth curve was defined from the logarithm of the geometric mean absolute error,
\begin{equation}
g(h)=\log\left[\operatorname{GM}_i\left(\max\{|\eta_i(h)|,10^{-12}\}\right)\right].
\end{equation}
Linear, piecewise-linear, and quadratic fits to $g(h)$ provided slope, curvature, fit-quality, residual roughness, monotonicity, stability, and diagnostic descriptors.

FEDE was calculated from the signed-error distribution at each horizon. Using $B=32$ fixed bins over a record-level pooled error range, the normalized entropy was
\begin{equation}
E(h)=-\frac{1}{\log B}\sum_{b:p_b(h)>0}p_b(h)\log p_b(h),
\end{equation}
where $p_b(h)$ is the proportion of signed errors in bin $b$. Horizon-specific values and summaries of the FEDE profile, including its early slope, intercept, range, and mean, were added to the feature set. Complete parameters and descriptor definitions are provided in Supplementary Methods; the implementation code is available from the authors upon reasonable request.

\subsection{Trial aggregation and feature nomenclature}

For each anatomical signal, descriptors were retained separately by side (SIDE) and combined as the bilateral mean (MEANLR), absolute asymmetry $|R-L|$ (LRABS), or signed asymmetry $R-L$ (LRSIGNED). The three trial-level values for a participant at each time point were averaged to obtain one subject-level feature profile. The same aggregation rule was used for continuous and indicator-type descriptors; missing trial values were omitted from the within-subject mean.

The primary between-group change analysis contained 1,098 analyzable features: 713 continuous and 385 discrete or indicator-type features. Two additional descriptors occurred in the within-cohort paired output but were absent from the between-group results and were not included in the primary feature count. To improve readability, the main text uses descriptive labels; the corresponding machine-readable identifiers are listed in Supplementary Table~S3.

\subsection{Primary longitudinal statistical analysis}

For participant $i$ and feature $f$, individual change was calculated as
\begin{equation}
\Delta f_i=f_{i,T1}-f_{i,T0}.
\end{equation}
Continuous change values were compared between NW and APA using one-way analysis of variance, with the $F$-ratio used for feature ranking and Cohen's $d$ used as a standardized effect size. Discrete or indicator-type changes were compared using the Mann--Whitney $U$ test with rank-biserial correlation (RBC). Group-specific change estimates and 95\% confidence intervals were obtained by bootstrap resampling. Benjamini--Hochberg FDR correction was performed separately for continuous and discrete feature families. Statistical significance required $q<0.05$.

Within each intervention group, T0 and T1 were additionally compared using paired $t$ tests for continuous features and Wilcoxon signed-rank tests for discrete features. These tests described within-group change but were not used to infer a difference between interventions: significance in one group and non-significance in another is not itself evidence that the groups changed differently.

\subsection{Baseline-adjusted sensitivity and technical checks}

Features with nominal between-group differences ($p<0.05$ before FDR correction) were subjected to baseline-adjusted sensitivity analysis using
\begin{equation}
T1=\beta_0+\beta_1T0+\beta_2\mathrm{Group}+\varepsilon.
\end{equation}
The coefficient for Group evaluated whether the T1 difference persisted after accounting for the baseline value of the same feature. Record length, effective forecast horizon, and walking speed were also compared at baseline and as change scores to evaluate potential technical confounding.

\subsection{Nested machine learning on individual change vectors}

To test whether the complete longitudinal response pattern identified the intervention, the vector $\Delta\mathbf{x}_i=\mathbf{x}_{i,T1}-\mathbf{x}_{i,T0}$ was used as the sole model input. A subject-wise nested pipeline comprised five repetitions of four outer folds and three inner folds. Feature ranking, feature-number selection, imputation, scaling, model selection, and hyperparameter optimization were confined to the inner training data. Candidate models included support-vector machines, random forests, and logistic regression. Performance was evaluated by collecting the complete out-of-fold predictions for all 24 participants within each four-fold partition and calculating one MCC for each repeat. Five repeated partitions therefore produced five complete-OOF MCC values, summarized by their mean and standard deviation. A separate 1,000-iteration full-pipeline permutation test used the mean four-fold cross-validation MCC as its test statistic and provided an empirical upper-tail $p$-value for positive above-chance classification.

\subsection{Exploratory cross-sectional cohort classification}

For comparison with the original analysis, NW-versus-APA classification was retained separately at T0 and T1. Single-feature thresholds and compact one- to three-feature models were searched using subject-isolated repeated validation. Because these analyses selected favorable features and combinations from large candidate pools, their peak MCC values are treated as internal exploratory estimates of cohort separability, not as independent evidence of rehabilitation effects. Detailed cross-sectional methods and results are reported only in the Supplementary Material.

\section{Results}
\label{sec:results}

\subsection{Primary comparison of individual change}

Eight of 713 continuous features and nine of 385 discrete features had nominal $p<0.05$. None met the FDR threshold: the minimum adjusted values were $q=0.977$ for continuous features and $q=0.930$ for discrete features. Thus, the data did not provide multiplicity-controlled evidence that any individual FEG-Pro/FEDE descriptor changed differently after NW and APA.

\begin{figure}[H]
\centering
\includegraphics[width=0.92\textwidth]{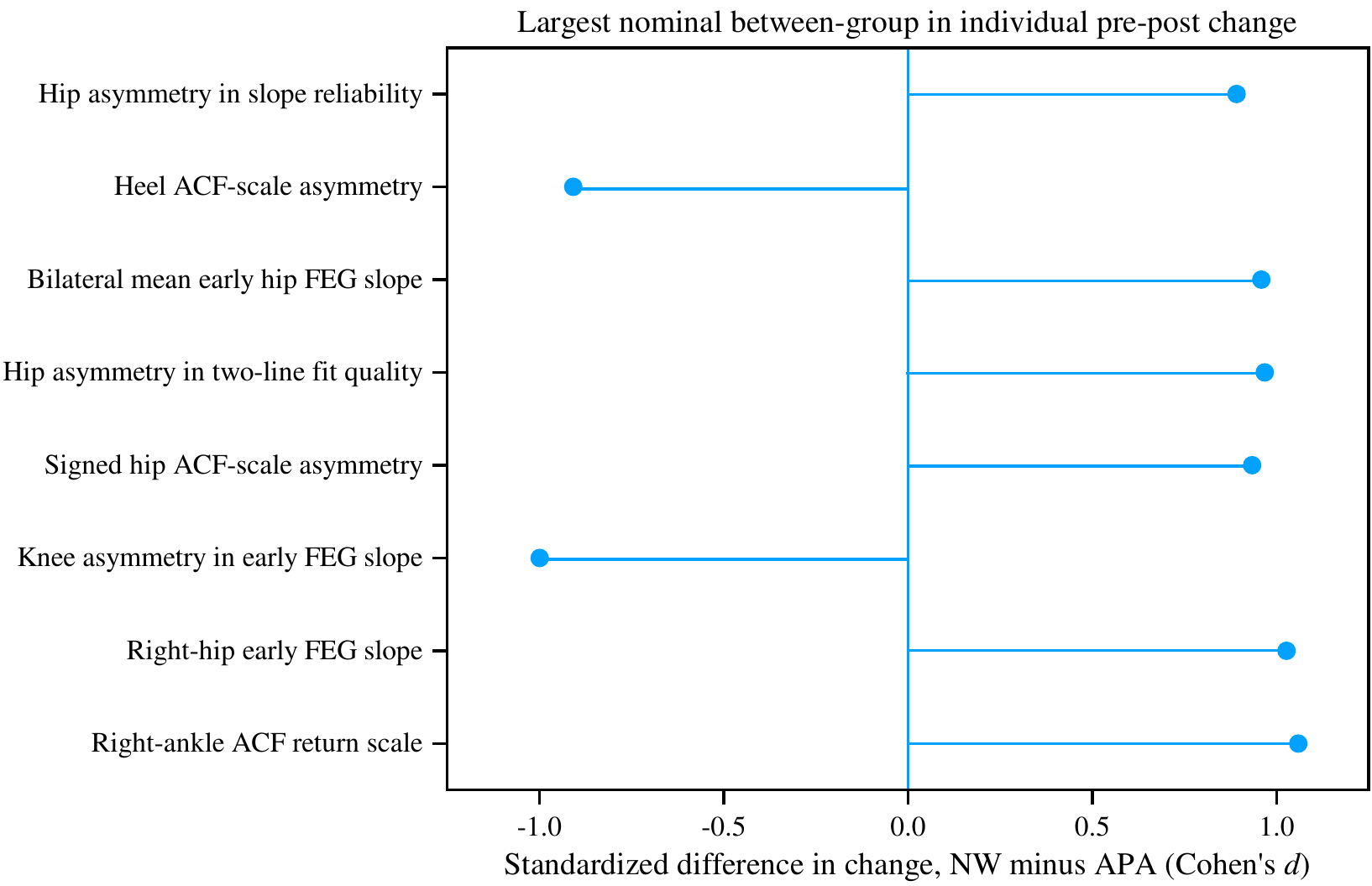}
\caption{Largest nominal differences between NW and APA in individual pre--post change for continuous features. Points show Cohen's $d$ for $\Delta_{NW} - \Delta_{APA}$; positive values indicate a larger or more positive change in NW. All displayed features had uncorrected $p < 0.05$, but none survived FDR correction (all $q = 0.977$). The horizontal segments are visual stems from zero and do not represent confidence intervals.}
\label{fig:candidates}
\end{figure}

The largest nominal continuous effect was the right-ankle autocorrelation return scale: it increased in NW ($\Delta=+4.738$) and decreased in APA ($\Delta=-11.833$), with $d=1.060$, $F=6.559$, and uncorrected $p=0.018$. Other nominal candidates involved early FEG slopes at the hip, bilateral knee-slope asymmetry, hip fit quality, and heel autocorrelation-scale asymmetry (Table~\ref{tab:candidates}; Figure~\ref{fig:candidates}). Indicator-type candidates included heel fit-regime descriptors and a left-knee negative-slope indicator. These patterns are useful for hypothesis generation but cannot be interpreted as confirmed intervention-specific effects.

\begin{table}[H]
\centering
\caption{Representative nominal candidate differences in pre--post change.}
\label{tab:candidates}
\begin{threeparttable}
\small
\begin{tabularx}{\textwidth}{>{\raggedright\arraybackslash}Xrrrr}
\toprule
Readable descriptor & NW change & APA change & Effect & $p$ / $q$ \\
\midrule
Right-ankle ACF return scale & $+4.738$ & $-11.833$ & $d=1.060$ & 0.018 / 0.977 \\
Right-hip early FEG slope & $+2.300$ & $-6.481$ & $d=1.028$ & 0.026 / 0.977 \\
Knee asymmetry in early FEG slope & $-1.959$ & $+3.274$ & $d=-0.998$ & 0.030 / 0.977 \\
Hip asymmetry in two-line fit quality & $+0.085$ & $-0.089$ & $d=0.969$ & 0.035 / 0.977 \\
Bilateral mean early hip FEG slope & $+2.039$ & $-3.819$ & $d=0.959$ & 0.037 / 0.977 \\
Heel ACF-scale asymmetry & $-0.381$ & $+0.367$ & $d=-0.907$ & 0.039 / 0.977 \\
Signed heel weak/mixed-fit asymmetry & $-0.167$ & $+0.167$ & RBC $=-0.550$ & 0.021 / 0.930 \\
Heel usable-linear-fit asymmetry & $+0.333$ & $-0.333$ & RBC $=0.507$ & 0.033 / 0.930 \\
Left-knee negative-slope indicator & $+0.167$ & $0.000$ & RBC $=0.471$ & 0.039 / 0.930 \\
\bottomrule
\end{tabularx}
\begin{tablenotes}[flushleft]\footnotesize
\item The table intentionally reports a compact, non-duplicated subset. Full machine-readable feature names, bootstrap intervals, test statistics, and complete feature-level results are available from the authors upon reasonable request. ACF, autocorrelation function; RBC, rank-biserial correlation.
\end{tablenotes}
\end{threeparttable}
\end{table}

\subsection{Within-group and baseline-adjusted analyses}

Separate paired analyses produced nominal pre--post findings within both programs, but none survived FDR correction. In the NW group, 25 features had uncorrected $p<0.05$ and the minimum adjusted value was approximately $q=0.733$; in the APA group, 41 features had uncorrected $p<0.05$ and the minimum adjusted value was approximately $q=0.671$. Therefore, these within-group rankings should also be regarded as exploratory.

Baseline-adjusted analyses were performed for all nominal between-group candidates. None retained a significant Group effect after adjustment for the baseline value of the same feature; the smallest $p$ was 0.060. The nominal effects were therefore attenuated after baseline adjustment. This result is compatible with baseline imbalance and low statistical power, but does not by itself prove that all candidate differences were caused by confounding. Record length and effective prediction horizon did not show significant differential changes, while baseline walking speed remained different between cohorts (Supplementary Table~S5).

\subsection{Machine learning on multidimensional change vectors}

The nested change-vector analysis failed to identify the intervention above chance. Across five repeated four-fold partitions, the complete out-of-fold predictions yielded a mean MCC of $-0.178\pm0.228$. In the separate 1,000-iteration permutation analysis, the observed mean four-fold MCC was $-0.601$, with an upper-tail $p=1.000$ for positive above-chance classification (two-sided $p=0.004$). The negative values indicate unstable or directionally inverted predictions in this small dataset and should not be interpreted as a biological “anti-learning” phenomenon.

\begin{figure}[H]
\centering
\includegraphics[width=0.82\textwidth]{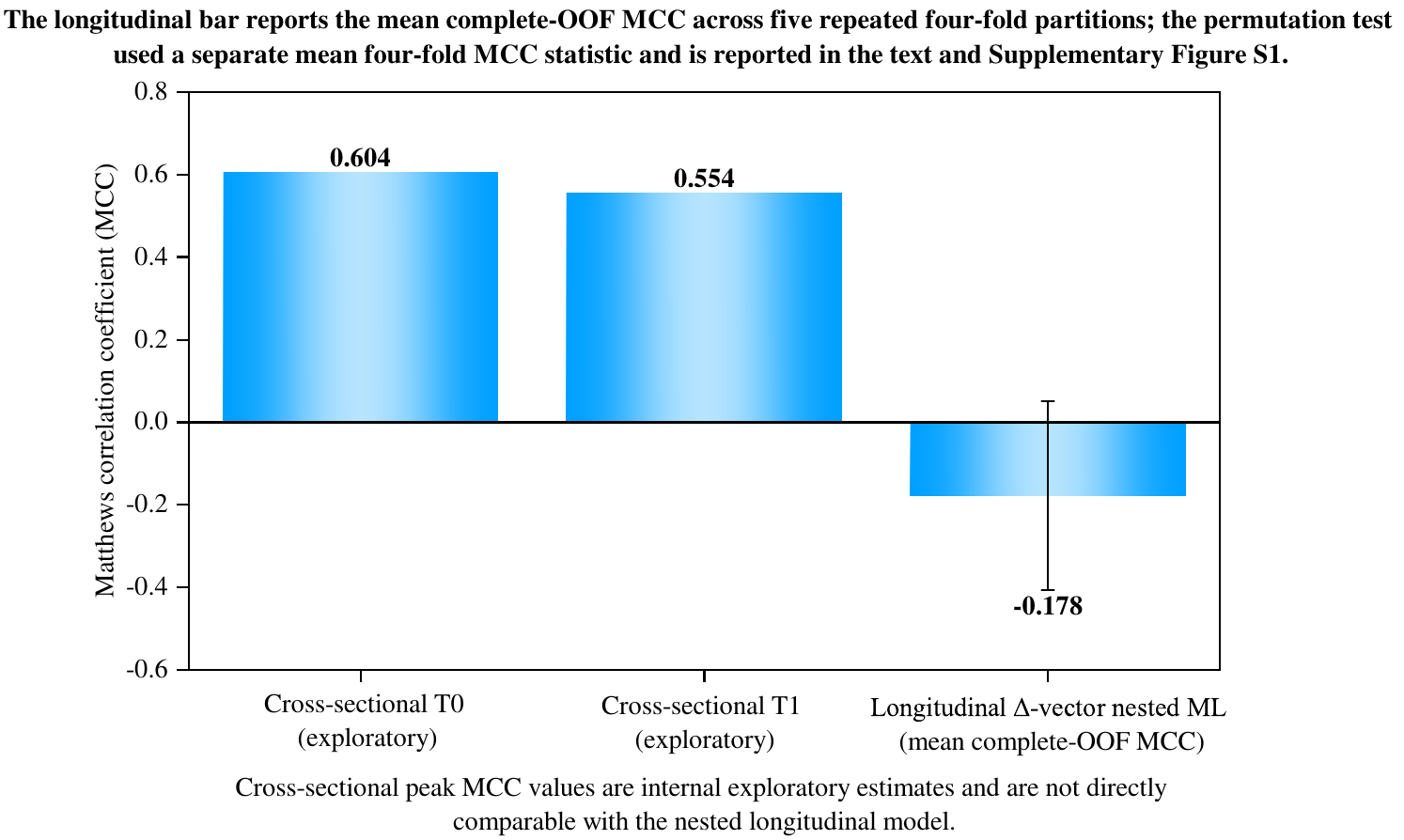}
\caption{Contrast between exploratory cross-sectional cohort separability and the primary longitudinal change-vector analysis. The cross-sectional peak MCC values were obtained after feature and model searches and are not directly comparable with the fully nested longitudinal estimate. Their coexistence with failed change-vector classification illustrates why cohort separability should not be interpreted as an intervention response.}
\label{fig:contrast}
\end{figure}

\subsection{Exploratory cross-sectional context}

The original cross-sectional searches produced peak internal MCC values of 0.604 at T0 and 0.554 at T1. At baseline, the highest-ranked features were dominated by hip-centered FEDE descriptors; at T1, knee- and heel-related FEG curve-shape and fit-reliability descriptors were more prominent. This ranking shift is descriptively interesting, but the primary longitudinal analysis did not establish that it represented a modality-specific rehabilitation effect. Full rankings, compact feature combinations, and descriptive figures are presented in Supplementary Tables~S6--S8 and Figures~S2--S4.

\section{Discussion}
\label{sec:discussion}

The principal finding is methodological as well as empirical. FEG-Pro generated a large set of nonlinear descriptors from short gait records, and exploratory cross-sectional models could separate the self-selected NW and APA cohorts both before and after the intervention. However, when the inferential target was changed to the individual longitudinal response, no feature survived multiplicity correction, baseline-adjusted candidate effects were non-significant, and nested machine learning could not identify the intervention from multidimensional change vectors. Cross-sectional separability and longitudinal response were therefore not equivalent in this dataset.

This distinction matters because the literature provides ample evidence that rehabilitation can modify gait in PD, including pace, rhythm, variability, balance, dual-task performance, joint biomechanics, cue-dependent coordination, muscle activation, and neural-control correlates \cite{ref12,ref13,ref14,ref15,ref16,ref17,ref18,ref19,ref20,ref21,ref22,ref23,ref24,ref25,ref26,ref27}. The absence of a robust NW-versus-APA difference here does not imply that neither intervention changed gait or provided clinical benefit. It means that, in a sample of 24 non-randomized participants, the available FEG-Pro descriptors did not demonstrate a reproducible difference between the longitudinal responses to the two active programs.

Several localized descriptors nevertheless deserve future study. The right-ankle autocorrelation return scale and early hip FEG slopes changed in opposite directions between groups and had standardized effects near or above one before multiplicity correction. Heel fit-regime indicators and knee slope/asymmetry features also appeared among nominal candidates. Such descriptors may reflect differences in temporal organization, predictability, or joint-specific adaptation. Their clinical direction cannot yet be labeled beneficial or adverse because the study lacked a healthy reference range and because the candidate effects were not statistically robust after FDR correction and baseline adjustment.

The work also clarifies the role of FEG-Pro. Its value is not established by obtaining a high classifier score after extensive feature search. Rather, FEG-Pro provides a structured family of short-record descriptors that can be subjected to clinically appropriate longitudinal validation. The framework captures complementary information about forecast-error growth geometry and uncertainty, and the present analysis illustrates a validation sequence that should accompany candidate digital biomarkers: subject-level aggregation, change-score analysis, effect sizes and confidence intervals, FDR control, baseline adjustment, nested modeling, and full-pipeline permutation testing.

The exploratory baseline and post-intervention classifiers remain useful as descriptive analyses of cohort structure. Their feature-family shift from hip/FEDE to knee/heel FEG descriptors may motivate targeted hypotheses. However, the baseline MCC of 0.604 arose before either program was delivered. It may therefore reflect self-selection, baseline physical capacity, latent motor phenotype, or model-selection optimism. The similar post-intervention MCC does not prove that an underlying predisposition was preserved, nor does the changed ranking prove that training created new biomarkers. Those interpretations require replication with randomized allocation, adequate power, and explicit modeling of Group-by-Time effects.

\section{Limitations}
\label{sec:limitations}

The study has several important limitations. First, the sample was small and the intervention groups were self-selected, with nominal baseline differences in clinical and gait variables. Second, the large feature pool created a severe multiple-testing burden; candidate effects with large uncorrected effect sizes remained highly uncertain. Third, the study did not include a healthy reference group, preventing interpretation of feature direction as normalization or deterioration. Fourth, trial-level discrete indicators were averaged across three trials, producing proportions that are convenient for subject-level analysis but require cautious physiological interpretation. Fifth, the cross-sectional feature/model search was exploratory and may be optimistic despite subject-isolated validation. Finally, the FEG-Pro configuration was fixed a priori; independent datasets are required to evaluate parameter sensitivity, test--retest reliability, and clinical validity.

\section{Conclusion}
\label{sec:conclusion}

FEG-Pro enabled nonlinear analysis of short hip, knee, ankle, and heel time series and identified localized candidate patterns of pre--post gait change. Nevertheless, this small, non-randomized cohort did not provide robust evidence that NW and APA produced different longitudinal FEG-Pro/FEDE responses. The central result is that cross-sectional machine-learning separability of rehabilitation cohorts---even when observed both before and after treatment---does not establish a differential intervention effect. Validation of rehabilitation-related nonlinear gait markers should be based on individual longitudinal change, baseline-aware statistics, multiplicity correction, and fully nested model evaluation.

\section*{Data availability}
The gait data are publicly available through the dataset published by Viglialoro et al. \cite{ref30}. The complete statistical result tables used in the revised analysis are available from the authors upon reasonable request.

\section*{Code availability}
The FEG-Pro feature-extraction code and the scripts used for the longitudinal statistical analyses, FDR correction, within-group paired analyses, baseline-adjusted sensitivity analysis, technical checks, nested machine-learning evaluation, and permutation testing are available from the authors upon reasonable request.

\section*{Ethics statement}
The original data collection was approved by the University of Pisa bioethics committee (No. 58/2024). Participants provided informed consent, and the protocol followed the Declaration of Helsinki. The present study is a secondary analysis of de-identified public data.

\section*{CRediT authorship contribution statement}
\textbf{Xuanbao Xiang:} Investigation, Data curation, Formal analysis, Writing -- original draft. \textbf{Andrei Velichko:} Methodology, Software, Formal analysis, Validation, Writing -- review and editing. \textbf{Xiaobo Rao:} Methodology, Supervision, Writing -- review and editing. \textbf{Jianshe Gao:} Conceptualization, Supervision, Writing -- review and editing.

\section*{Declaration of competing interest}
The authors declare that they have no known competing financial interests or personal relationships that could have appeared to influence the work reported in this paper.

\section*{Funding}
This research did not receive any specific grant from funding agencies in the public, commercial, or not-for-profit sectors.

\clearpage


\begin{center}
\LARGE\textbf{Supplementary Material}
\end{center}
\vspace{1em}

\setcounter{table}{0}
\setcounter{figure}{0}
\setcounter{section}{0}
\renewcommand{\thetable}{S\arabic{table}}
\renewcommand{\thefigure}{S\arabic{figure}}

\section*{Supplementary Methods}

\subsection*{FEG-Pro parameter configuration}

\begin{table}[H]
\centering
\caption{Fixed FEG-Pro parameters used in the gait analysis.}
\label{tab:S1}
\small
\begin{tabularx}{\textwidth}{p{4.4cm}X}
\toprule
Parameter & Value \\
\midrule
Sampling rate & 100 Hz \\
Internal resampling & Disabled \\
History span & $H=300$ samples \\
Sparse embedding dimension & $m=30$ \\
Forecast horizons & $h=1,2,\ldots,14$ \\
KNN neighbors & $k=3$, distance weighted \\
Temporal split within each scalar record & 60\% training / 40\% testing \\
Feature standardization & Training-set z-score \\
FEDE histogram bins & 32 \\
FEDE histogram range & Pooled signed-error range across horizons within each record \\
\bottomrule
\end{tabularx}
\end{table}

The sparse history-vector offsets were distributed approximately uniformly over the interval $0,\ldots,H-1$. The feature-extraction code limited the usable forecast horizon according to record length and omitted horizons without sufficient training/test samples. No artificial interpolation or resampling was applied to the gait signals.

\subsection*{Feature naming and readable labels}

\begin{table}[H]
\centering
\caption{Feature-prefix definitions and readable labels used in the main manuscript.}
\label{tab:S2}
\small
\begin{tabularx}{\textwidth}{p{2.5cm}p{5.0cm}X}
\toprule
Prefix/metric & Meaning & Example/readable form \\
\midrule
\texttt{SIDE} & Side-specific descriptor & Right-ankle ACF return scale \\
\texttt{MEANLR} & Mean of left and right descriptors & Bilateral mean early hip FEG slope \\
\texttt{LRABS} & Absolute left--right difference $|R-L|$ & Knee asymmetry in early FEG slope \\
\texttt{LRSIGNED} & Signed difference $R-L$ & Signed hip ACF-scale asymmetry \\
\texttt{zpm} & First positive-to-negative ACF zero crossing & ACF zero-crossing scale \\
\texttt{zmp} & Subsequent negative-to-positive ACF return & ACF return scale \\
\bottomrule
\end{tabularx}
\end{table}

The remaining metric names describe the fitted FEG curve or its diagnostic regime. For example, \texttt{slope\_two\_left} denotes the early segment of a two-line fit, \texttt{R2\_two\_min} denotes the minimum local fit quality of that model, and \texttt{lambda\_reliability\_score} summarizes the stability of the selected FEG slope. Indicator names beginning with \texttt{diagnosis\_} encode the selected curve-shape regime.

\begin{table}[H]
\centering
\caption{Machine-readable identifiers for the candidate features shown in the main manuscript.}
\label{tab:S3}
\footnotesize
\begin{tabularx}{\textwidth}{p{5.0cm}X}
\toprule
Readable label & Machine-readable identifier \\
\midrule
Right-ankle ACF return scale & \path{SIDE__zmp_used__R__ankle} \\
Right-hip early FEG slope & \path{SIDE__slope_two_left__R__hip} \\
Knee asymmetry in early FEG slope & \path{LRABS__slope_two_left__knee} \\
Hip asymmetry in two-line fit quality & \path{LRABS__R2_two_min__hip} \\
Signed hip ACF-scale asymmetry & \path{LRSIGNED_RminusL__zpm_used__hip} \\
Bilateral mean early hip FEG slope & \path{MEANLR__slope_two_left__hip} \\
Heel ACF-scale asymmetry & \path{LRABS__zpm_used__heel} \\
Hip asymmetry in slope reliability & \path{LRABS__lambda_reliability_score__hip} \\
Signed heel weak/mixed-fit asymmetry & \path{LRSIGNED_RminusL__diagnosis_weak_or_mixed__heel} \\
Heel usable-linear-fit asymmetry & \path{LRABS__diagnosis_usable_linear__heel} \\
Left-knee negative-slope indicator & \path{SIDE__lambda_sign_negative__L__knee} \\
\bottomrule
\end{tabularx}
\end{table}

\section*{Supplementary Figures}

\begin{figure}[H]
\centering
\includegraphics[width=\textwidth]{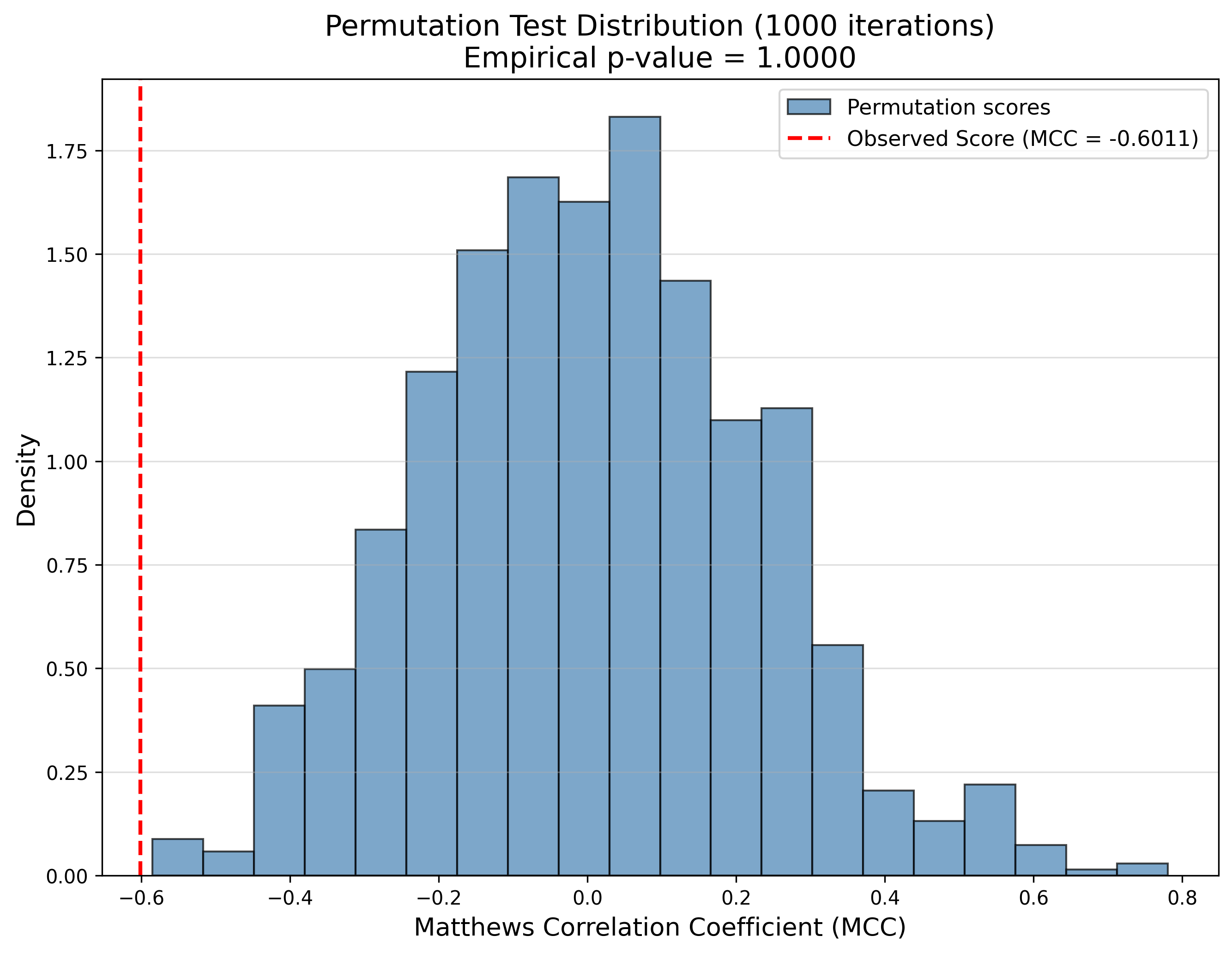}
\caption{Null distribution from the 1,000-iteration full-pipeline permutation test for intervention classification using multidimensional change vectors. The vertical line shows the observed mean four-fold cross-validation MCC ($-0.6011$) used as the test statistic in the 1,000-iteration full-pipeline permutation procedure. This value is distinct from the mean complete-OOF MCC calculated across the five repeated partitions in the primary evaluation. The result indicates failure to demonstrate above-chance intervention classification and unstable directionally inverted predictions in this small sample.}
\label{fig:S1}
\end{figure}

\begin{figure}[H]
\centering
\includegraphics[width=0.82\textwidth]{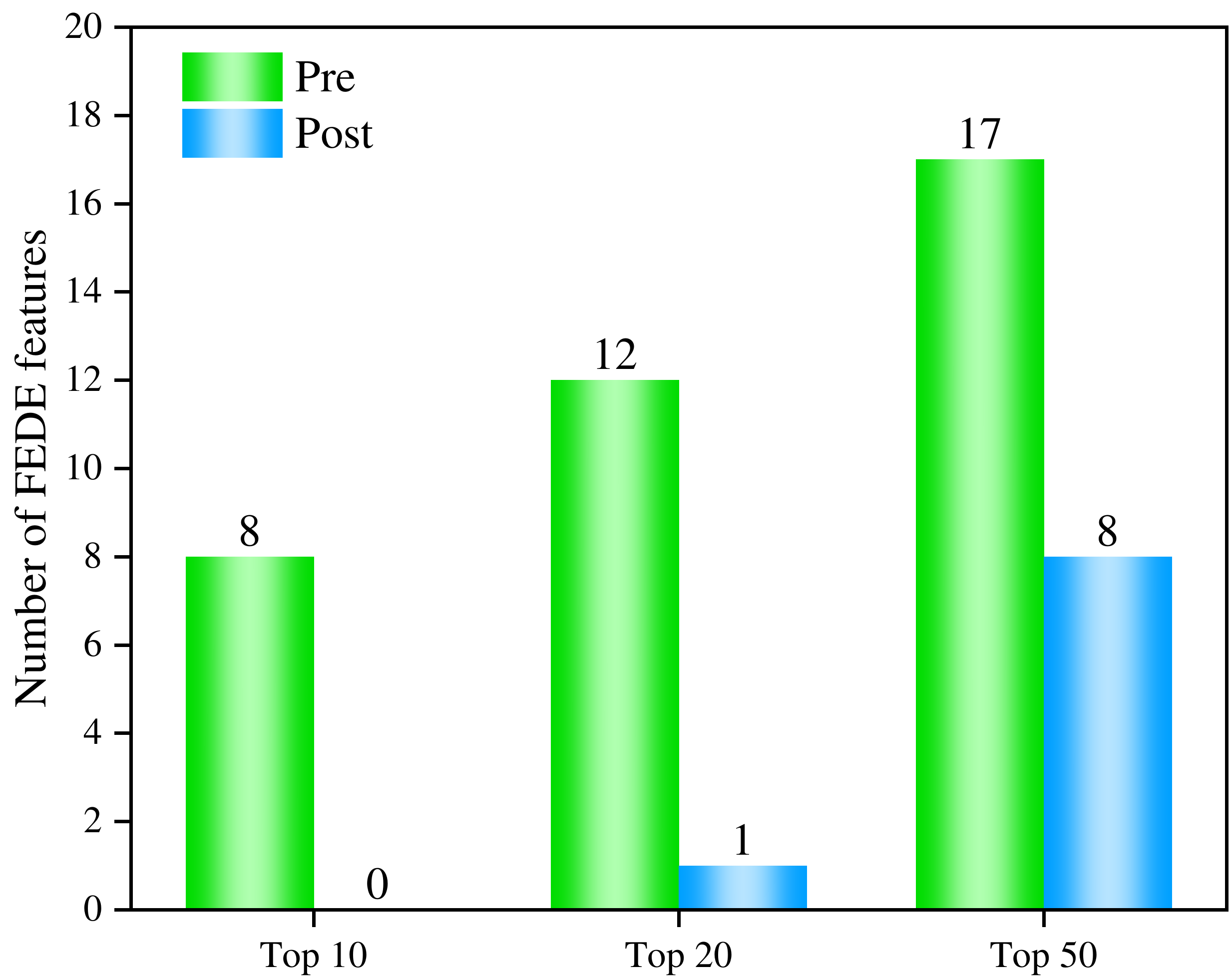}
\caption{Exploratory changes in the number of FEDE descriptors among the top-ranked cross-sectional single features at T0 and T1. These rankings describe cohort separability and are not interpreted as intervention effects.}
\label{fig:S2}
\end{figure}

\begin{figure}[H]
\centering
\includegraphics[width=0.82\textwidth]{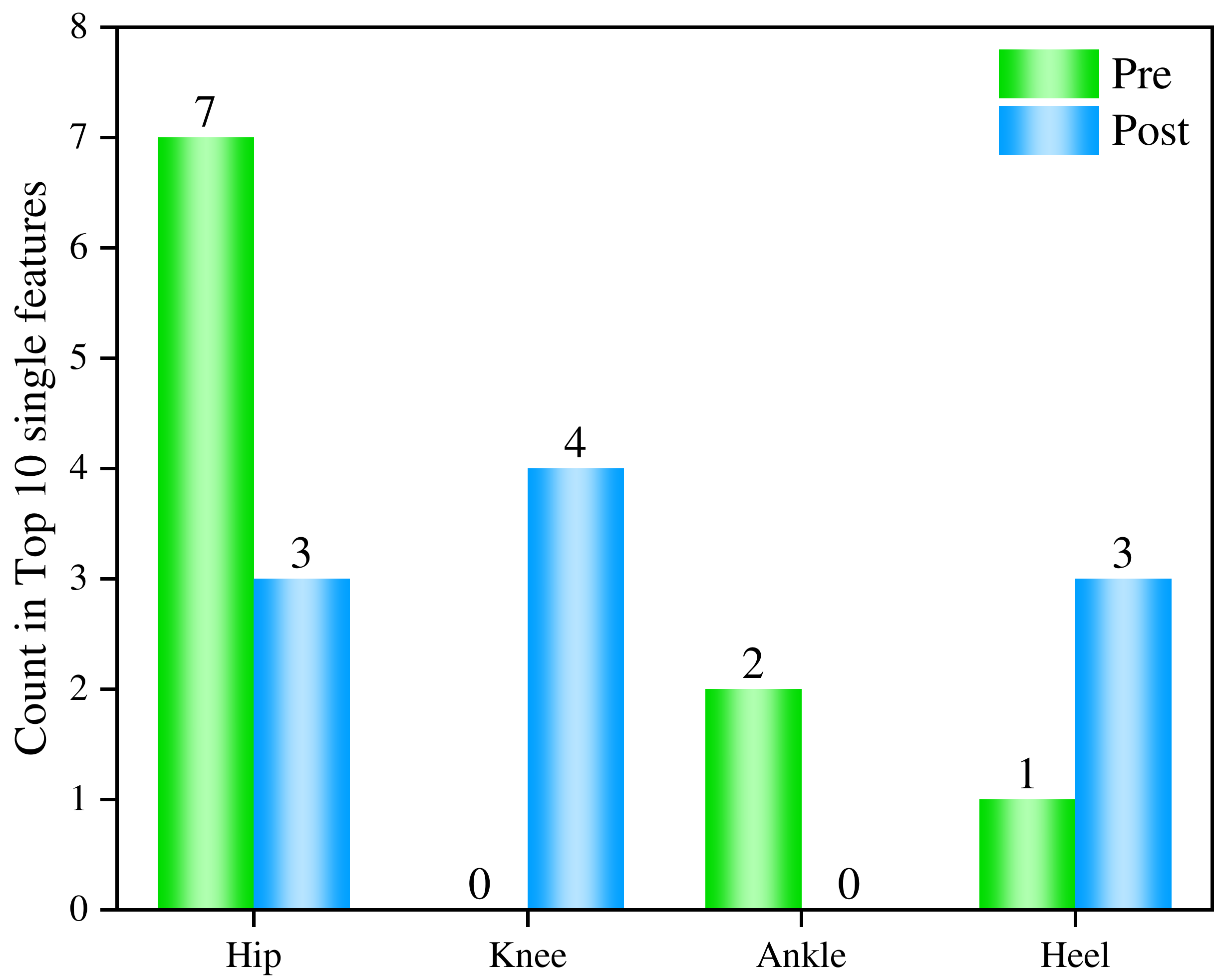}
\caption{Exploratory anatomical distribution of the top 10 cross-sectional single features at T0 and T1.}
\label{fig:S3}
\end{figure}

\begin{figure}[H]
\centering
\includegraphics[width=0.82\textwidth]{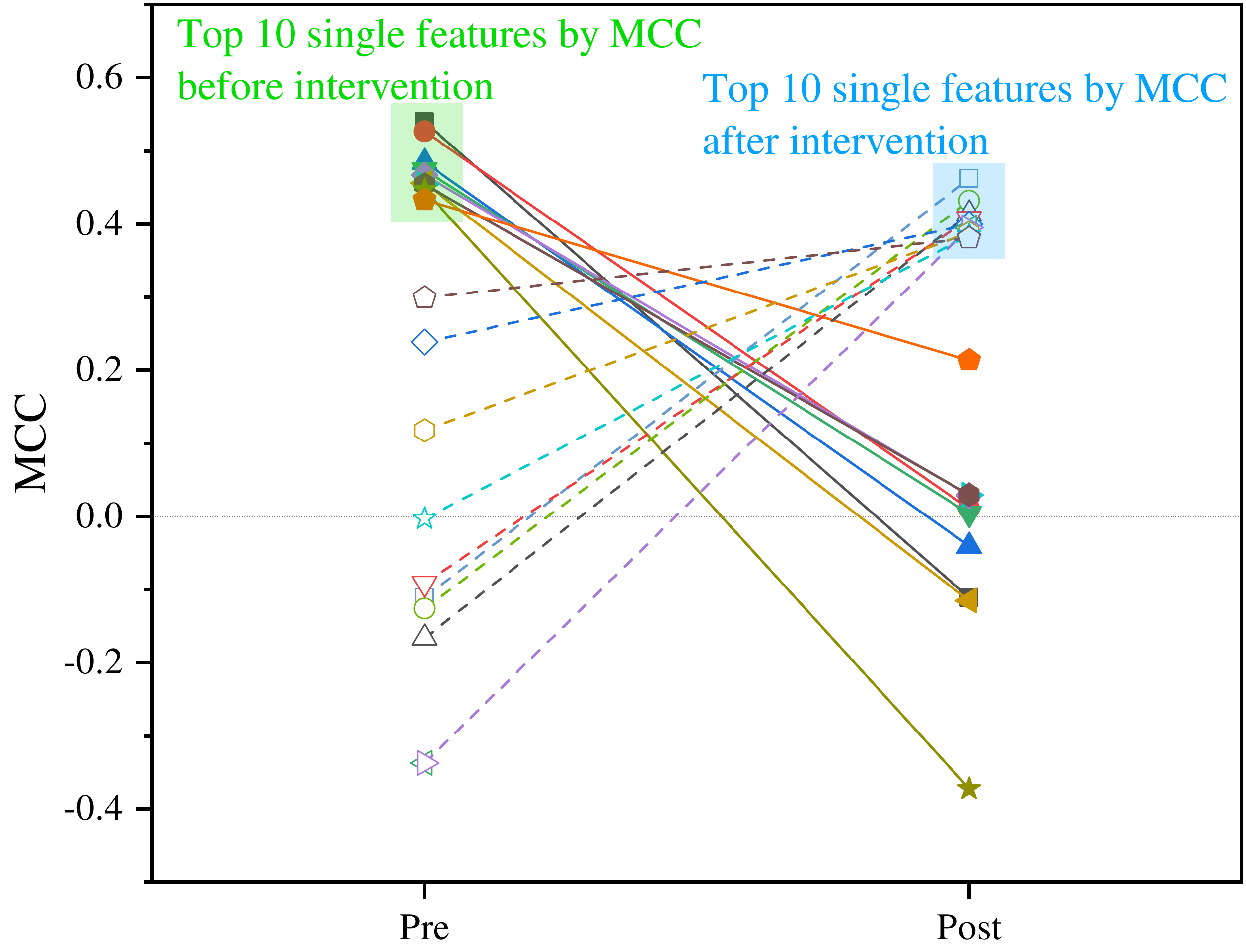}
\caption{Exploratory trajectories of cross-sectional feature MCC values from T0 to T1. These internal estimates were obtained after feature screening and are presented as descriptive context only. Solid lines indicate how the discriminative power of these initially dominant features diminished following the 12-week intervention. Dashed lines illustrate that these newly emerged discriminative features lacked separability at baseline but exhibited increases in MCC after the intervention.}
\label{fig:S4}
\end{figure}

\clearpage
\section*{Supplementary Tables}

\begin{table}[H]
\centering
\caption{Baseline-adjusted sensitivity analysis for nominal candidate features.}
\label{tab:S4}
\resizebox{\textwidth}{!}{
\begin{tabular}{lcccccc}
\toprule
Feature & T0 Coefficient & 95\% CI (T0) & p (T0) & Group Coefficient & 95\% CI (Group) & p (Group) \\
\midrule
\texttt{LRABS\_\allowbreak\_zpm\_\allowbreak\_used\_\allowbreak\_heel} & 0.504 & [0.127, 0.881] & 0.011 & -0.606 & [-1.239, 0.027] & 0.060 \\
\texttt{MEANLR\_\allowbreak\_slope\_\allowbreak\_two\_\allowbreak\_left\_\allowbreak\_hip} & -0.058 & [-0.695, 0.579] & 0.851 & 3.690 & [-0.888, 8.268] & 0.108 \\
\texttt{SIDE\_\allowbreak\_slope\_\allowbreak\_two\_\allowbreak\_left\_\allowbreak\_R\_\allowbreak\_hip} & 0.085 & [-0.531, 0.700] & 0.776 & 5.433 & [-1.344, 12.210] & 0.110 \\
\texttt{LRABS\_\allowbreak\_diagnosis\_\allowbreak\_usable\_\allowbreak\_linear\_\allowbreak\_heel} & 0.200 & [-0.296, 0.695] & 0.412 & 0.194 & [-0.049, 0.437] & 0.111 \\
\texttt{SIDE\_\allowbreak\_lambda\_\allowbreak\_sign\_\allowbreak\_negative\_\allowbreak\_L\_\allowbreak\_knee} & 0.197 & [-0.291, 0.684] & 0.411 & 0.152 & [-0.045, 0.350] & 0.124 \\
\texttt{LRABS\_\allowbreak\_slope\_\allowbreak\_two\_\allowbreak\_left\_\allowbreak\_knee} & -0.362 & [-1.131, 0.407] & 0.337 & -2.826 & [-6.741, 1.090] & 0.147 \\
\texttt{LRSIGNED\_\allowbreak\_RminusL\_\allowbreak\_zpm\_\allowbreak\_used\_\allowbreak\_hip} & 0.296 & [-0.176, 0.767] & 0.206 & 0.315 & [-0.246, 0.877] & 0.256 \\
\texttt{SIDE\_\allowbreak\_zmp\_\allowbreak\_used\_\allowbreak\_R\_\allowbreak\_ankle} & 0.455 & [0.149, 0.761] & 0.006 & 6.382 & [-5.766, 18.529] & 0.287 \\
\texttt{SIDE\_\allowbreak\_diagnosis\_\allowbreak\_curved\_\allowbreak\_growth\_\allowbreak\_R\_\allowbreak\_ankle} & 0.238 & [-0.075, 0.552] & 0.129 & -0.066 & [-0.200, 0.068] & 0.318 \\
\texttt{SIDE\_\allowbreak\_best\_\allowbreak\_r2\_\allowbreak\_model\_\allowbreak\_quadratic\_\allowbreak\_R\_\allowbreak\_ankle} & 0.238 & [-0.075, 0.552] & 0.129 & -0.066 & [-0.200, 0.068] & 0.318 \\
\texttt{SIDE\_\allowbreak\_diagnosis\_\allowbreak\_weak\_\allowbreak\_or\_\allowbreak\_mixed\_\allowbreak\_R\_\allowbreak\_heel} & 0.203 & [-0.420, 0.827] & 0.506 & -0.130 & [-0.400, 0.140] & 0.327 \\
\texttt{LRABS\_\allowbreak\_R2\_\allowbreak\_two\_\allowbreak\_min\_\allowbreak\_hip} & -0.036 & [-0.762, 0.689] & 0.917 & 0.060 & [-0.097, 0.218] & 0.434 \\
\texttt{LRABS\_\allowbreak\_lambda\_\allowbreak\_reliability\_\allowbreak\_score\_\allowbreak\_hip} & 0.352 & [-0.345, 1.050] & 0.306 & 0.041 & [-0.068, 0.150] & 0.442 \\
\texttt{LRABS\_\allowbreak\_diagnosis\_\allowbreak\_curved\_\allowbreak\_growth\_\allowbreak\_ankle} & -0.127 & [-0.492, 0.238] & 0.477 & 0.030 & [-0.113, 0.172] & 0.670 \\
\texttt{LRABS\_\allowbreak\_best\_\allowbreak\_r2\_\allowbreak\_model\_\allowbreak\_two\_\allowbreak\_lines\_\allowbreak\_ankle} & -0.127 & [-0.492, 0.238] & 0.477 & 0.030 & [-0.113, 0.172] & 0.670 \\
\texttt{LRABS\_\allowbreak\_best\_\allowbreak\_r2\_\allowbreak\_model\_\allowbreak\_quadratic\_\allowbreak\_ankle} & -0.127 & [-0.492, 0.238] & 0.477 & 0.030 & [-0.113, 0.172] & 0.670 \\
\texttt{LRSIGNED\_\allowbreak\_RminusL\_\allowbreak\_diagnosis\_\allowbreak\_weak\_\allowbreak\_or\_\allowbreak\_mixed\_\allowbreak\_heel} & 0.220 & [-0.177, 0.616] & 0.262 & -0.057 & [-0.335, 0.221] & 0.673 \\
\bottomrule
\end{tabular}
}
\begin{flushleft}
\footnotesize{No candidate retained $p<0.05$ after adjustment for its baseline value. Technical input/used duplicates are omitted from this compact table.}
\end{flushleft}
\end{table}

\begin{table}[H]
\centering
\caption{Record length, effective horizon, and walking-speed checks.}
\label{tab:S5}
\resizebox{\textwidth}{!}{
\begin{tabular}{llccccc}
\toprule
Variable & Cohort & T0 & T1 & Within-group $p$ & Between-group $p$ (T0) & Between-group $p$ ($\Delta$) \\
\midrule
Record length & NW & $435.45\pm95.22$ & $416.93\pm108.87$ & 0.567 & 0.238 & 0.713 \\
 & APA & $480.47\pm85.06$ & $444.30\pm119.79$ & 0.330 & & \\
Effective horizon & NW & 14.0 [14.0--14.0] & 14.0 [10.5--14.0] & 0.343 & 0.246 & 0.625 \\
 & APA & 14.0 [14.0--14.0] & 14.0 [8.5--14.0] & 0.250 & & \\
Walking speed & NW & $0.960\pm0.214$ & $1.003\pm0.221$ & 0.213 & 0.034 & 0.351 \\
 & APA & $0.794\pm0.146$ & $0.889\pm0.211$ & 0.060 & & \\
\bottomrule
\end{tabular}}
\end{table}

\begin{table}[H]
\centering
\caption{Top 10 exploratory cross-sectional features at baseline.}
\label{tab:S6}
\resizebox{\textwidth}{!}{
\begin{tabular}{rllccl}
\toprule
Rank & Feature & Level & MCC & Anatomy & Family \\
\midrule
1 & \texttt{LRABS\_\_fit\_confidence\_medium\_\_heel} & subject & 0.540 & heel & non-FEDE \\
2 & \texttt{MEANLR\_\_fede\_h1\_norm\_\_hip} & trial & 0.527 & hip & FEDE \\
3 & \texttt{MEANLR\_\_fede\_early\_intercept\_norm\_\_hip} & trial & 0.485 & hip & FEDE \\
4 & \texttt{MEANLR\_\_fede\_early\_intercept\_norm\_\_hip} & subject & 0.474 & hip & FEDE \\
5 & \texttt{MEANLR\_\_fede\_h1\_norm\_\_hip} & subject & 0.467 & hip & FEDE \\
6 & \texttt{SIDE\_\_fede\_min\_norm\_\_R\_\_hip} & subject & 0.456 & hip & FEDE \\
7 & \texttt{LRABS\_\_fede\_early\_slope\_\_ankle} & subject & 0.454 & ankle & FEDE \\
8 & \texttt{LRABS\_\_fede\_early\_slope\_norm\_\_ankle} & subject & 0.454 & ankle & FEDE \\
9 & \texttt{SIDE\_\_fede\_mean\_norm\_\_R\_\_hip} & subject & 0.447 & hip & FEDE \\
10 & \texttt{LRABS\_\_monotonicity\_fraction\_up\_\_hip} & subject & 0.433 & hip & non-FEDE \\
\bottomrule
\end{tabular}}
\end{table}

\begin{table}[H]
\centering
\caption{Top 10 exploratory cross-sectional features after intervention.}
\label{tab:S7}
\resizebox{\textwidth}{!}{
\begin{tabular}{rllccl}
\toprule
Rank & Feature & Level & MCC & Anatomy & Family \\
\midrule
1 & \texttt{SIDE\_\_lambda\_sign\_negative\_\_L\_\_knee} & subject & 0.462 & knee & non-FEDE \\
2 & \texttt{MEANLR\_\_curvature\_gain\_\_heel} & subject & 0.432 & heel & non-FEDE \\
3 & \texttt{SIDE\_\_lambda\_reliability\_high\_\_L\_\_knee} & subject & 0.414 & knee & non-FEDE \\
4 & \texttt{MEANLR\_\_curvature\_gain\_\_heel} & trial & 0.407 & heel & non-FEDE \\
5 & \texttt{LRABS\_\_diagnosis\_usable\_linear\_\_heel} & subject & 0.400 & heel & non-FEDE \\
6 & \texttt{SIDE\_\_fit\_confidence\_low\_\_L\_\_knee} & subject & 0.395 & knee & non-FEDE \\
7 & \texttt{SIDE\_\_lambda\_reliability\_medium\_\_L\_\_knee} & subject & 0.395 & knee & non-FEDE \\
8 & \texttt{LRABS\_\_diagnosis\_weak\_or\_mixed\_\_hip} & subject & 0.388 & hip & non-FEDE \\
9 & \texttt{MEANLR\_\_lambda\_reliability\_score\_\_hip} & subject & 0.385 & hip & non-FEDE \\
10 & \texttt{LRABS\_\_fit\_confidence\_medium\_\_hip} & subject & 0.381 & hip & non-FEDE \\
\bottomrule
\end{tabular}}
\end{table}

\begin{table}[H]
\centering
\caption{Best exploratory compact cross-sectional models.}
\label{tab:S8}
\resizebox{\textwidth}{!}{
\begin{tabular}{cllp{8.3cm}c}
\toprule
Time & Size & Model & Features & MCC \\
\midrule
T0 & single & KNN-3 & \texttt{MEANLR\_\_fede\_early\_intercept\_norm\_\_hip} & 0.432 \\
T0 & pair & KNN-3 & previous feature + \texttt{LRABS\_\_fit\_confidence\_medium\_\_heel} & 0.594 \\
T0 & triple & KNN-3 & previous pair + \texttt{MEANLR\_\_best\_r2\_model\_quadratic\_\_ankle} & 0.604 \\
T1 & single & linear SVM & \texttt{SIDE\_\_curvature\_gain\_\_L\_\_ankle} & 0.368 \\
T1 & pair & depth-2 tree & \texttt{MEANLR\_\_curvature\_gain\_\_heel} + \texttt{MEANLR\_\_lambda\_\_heel} & 0.520 \\
T1 & triple & depth-2 tree & previous pair + \texttt{LRSIGNED\_RminusL\_\_R2\_two\_local\_weighted\_\_knee} & 0.554 \\
\bottomrule
\end{tabular}}
\begin{flushleft}\footnotesize These peak internal estimates followed exhaustive feature-combination searches and are not independent validation results.\end{flushleft}
\end{table}

\end{document}